\documentclass{appolb}
\usepackage{graphicx}
\begin{document}
\title{PAIRING REENTRANCE IN \\ WARM ROTATING $^{104}$Pd NUCLEUS
}
\author{N. Quang Hung$^a$, \underline{N. Dinh Dang}$^b$, B.K. Agrawal$^{c}$, V.M. Datar$^{d}$, A. Mitra$^{d}$, and D. R. Chakrabarty$^{d}$
\address{$^a$ School of engineering, Tan Tao University, Tan Tao University Avenue, Tan Duc E. City, Duc Hoa, Long An Province, Vietnam \\
$^b$ Theoretical Nuclear Physics Laboratory, RIKEN Nishina Center for Accelerator-Based Science, 2-1 Hirosawa, Wako City, 351-0198 Saitama, Japan and Institute for Nuclear Science and Technique, Hanoi, Vietnam \\
$^c$ Theory Division, Saha Institute of Nuclear Physics, 1/AF, Bidhan Nagar, Kolkata 700 064, India \\
$^d$ Nuclear Physics Division, Bhabha Atomic Research Center, Mumbai 40085, India}
}
\maketitle
\begin{abstract}
Pairing reentrance phenomenon in the warm rotating $^{104}$Pd nucleus is studied within the Bardeen-Cooper-Schrieffer (BCS)-based approach (the FTBCS1). The theory takes into account the effect of quasiparticle number fluctuations on the pairing field at finite temperature and angular momentum within the pairing model plus noncollective rotation along the symmetry axis. The numerical calculations for the pairing gaps and nuclear level densities (NLD), of which an anomalous enhancement has been experimentally observed at low excitation energy $E^*$ and high angular momentum $J$, show that the pairing reentrance is seen in the behavior of pairing gap obtained within the FTBCS1 at low $E$ and high $J$. This leads to the enhancement of the FTBCS1 level densities, in good agreement with the experimental observation. This agreement indicates that the observed enhancement of the NLD might be the first experimental detection of the pairing reentrance in a finite nucleus.
\end{abstract}
\section{Introduction}
It is well-known that pairing correlation is strongly affected by both temperature $T$ and angular momentum $J$. The increase of temperature or angular momentum of a nucleus breaks the nucleon pairs located around the Fermi surface, which are responsible for the pairing. The nucleons from the broken pairs scatter to the single-particle levels nearby and completely block these levels. Consequently, pairing correlation decreases. When the temperature or angular momentum is high enough, i.e., equal to its critical value $T_c$ or $J_c$, these unpaired particles will block all the levels around the Fermi surface, completely destroying the pairing correlation. However, when both temperature and rotation are combined as in warm/hot rotating systems, there appears an interesting effect of pairing reentrance. When the angular momentum of the system is slightly higher than $J_c$, the pairing correlation, which is zero at low $T < T_1$, reappears at $T > T_1$, and decreases to vanish at $T_2 > T_1$. This phenomenon was first predicted by Kammuri \cite{Kam} by applying the Bardeen-Cooper-Schrieffer theory at finite temperature (FTBCS) and angular momentum to finite nuclear systems. It was later explained by Moretto \cite{Moretto} as follows. In the systems with $J > J_c$, the increase of temperature relaxes the unpaired nucleons located around the Fermi surface and therefore opens the possibilities for recreation of the pairs. However, if $T$ increases further, all of these newly created pairs will eventually be broken again, leading to the quenching of pairing correlation. A similar effect called unconventional superconductivity has been recently discovered in a superconducting URhGe material at Grenoble High Magnetic Field Laboratory \cite{Levy}. In this experiment, URhGe is normal when the magnetic field $H$, which plays the role as that of nuclear rotation, is around 2 Tesla, and becomes superconducting at low $T$ when the magnetic field increases between 8 and 13 Tesla.

It has been pointed out in several theoretical calculations that the collapse of pairing correlation at $T = T_c$ or $M = M_c$, which signals the superfluid-normal (SN) phase transition, is a shortcoming of the FTBCS theory because it neglects the thermal fluctuations in finite systems such as atomic nuclei. In nuclear systems, these thermal fluctuations are so large that they smooth out of the SN phase transition and lead to the non-collapsing of pairing correlation at $T \geq T_c$ \cite{Mang}. As the result, by taking into account the thermal fluctuations, the calculations within an exactly solvable model \cite{Frau} and the cranked shell model \cite{Sheikh} have shown a different behavior of the pairing reentrance phenomenon for which pairing gap, which is zero at $J > J_c$ and $T = 0$, reappears at a given $T$ but does not vanish as $T$ increases further. Similarly, by including the quasiparticle-number fluctuations (QNF) at finite temperature and angular momentum, the recent FTBCS1 theory has also discovered an identical behavior of the pairing reentrance phenomenon in several realistic nuclei \cite{AFTBCS1}. Apart from the FTBCS1, the shell model quantum Monte Carlo simulation for $^{72}$Ge nucleus has suggested that the pairing reentrance can be seen not only in the behavior of pairing gap but also in the local enhancement of the nuclear level density (NLD) at low $T$ and sufficiently high $J$~\cite{Dean}. This suggestion has given the feasibility for this phenomenon to be experimentally observed. In fact, the very recent series of experiments conducted at the Bhabha Atomic Research Center (BARC) for the reaction $^{12}$C + $^{93}$Nb $\rightarrow$ $^{105}$Ag$^*$ $\rightarrow$ $^{104}$Pd$^* + p$ at the incident energy of 40 - 45 MeV has observed an enhancement of the NLD of $^{104}$Pd nucleus at low excitation energy $E^*$ and high $J$, which is qualitatively similar to that predicted by the shell model Monte Carlo calculations~\cite{Mitra}. 

The aim of present study is to apply the FTBCS1 theory including finite angular momentum to study if the enhancement observed in the experimentally extracted NLD of $^{104}$Pd can be interpreted as the first evidence of pairing reentrance in a warm rotating finite nucleus.
\section{FTBCS1 theory}
The pairing Hamiltonian, which describes a spherical nucleus non-collectively rotating about the symmetry axis, chosen to coincide with its $z$ component, has the form 
\begin{equation}
H=\sum_k{\epsilon_k(a_{+k}^\dagger a_{+k} + a_{-k}^\dagger a_{-k}}) - G\sum_{kk'}{a_k^\dagger a_{-k}^\dagger a_{-k'} a_{k'}} - \lambda \hat{N} - \omega \hat{M} ~, 	\label{Hp}
\end{equation}
where $a_{\pm k}^\dagger (a_{\pm k})$ are the creation (annihilation) operators of a particle in the $k$-th deformed state, whereas $\epsilon_k, \lambda$, and $\omega$ are the single-particle energies, chemical potential, and rotational frequency, respectively. The particle-number operator $\hat{N}$ and the $z$-projection $\hat{M}$ of the total angular momentum $\hat{J}$ are defined as
\begin{equation}
\hat{N}=\sum_k(a_{+k}^{\dagger}a_{+k} + a_{-k}^{\dagger}a_{-k})~, \hspace{5mm}
\hat{M}=\sum_k m_k(a_{+k}^{\dagger}a_{+k} - a_{-k}^{\dagger}a_{-k}) ~, 	\label{NM}
\end{equation}
with the single-particle spin projection $m_k$.

The FTBCS1 equations including angular momentum are derived based on the variational procedures to minimize the expectation value of the pairing Hamiltonian (\ref{Hp}) in the grand-canonical ensembles \cite{Moretto}. The details are reported in Ref. \cite{AFTBCS1}, so we do not repeat it here. The final FTBCS1 equations for the pairing gap $\Delta$, particle number $N$, and total angular momentum $M$ are given as
\begin{equation}
\Delta_k = \Delta + \delta\Delta_k ~, \label{Gapeq}
\end{equation}
\begin{equation}
N = 2\sum_k\left[v_k^2(1-n_k^+ -n_k^{-}) + \frac{1}{2}(n_k^\dagger+n_k^{-}) \right]~,
\label{N}
\end{equation}
\begin{equation}
M = \sum_k m_k(n_k^+ - n_k^{-}) ~, \label{NM1}
\end{equation}
where 
\begin{equation}
\Delta=G\sum_{k'}{u_{k'}v_{k'}(1-n_{k'}^+-n_{k'}^-)}~, \hspace{5mm}
\delta\Delta_k=G\frac{\delta{\cal N}_k^2}{1-n_k^+-n_k^-}u_kv_k ~, \label{Gapeq1} 
\end{equation}
with $\delta{\cal N}_k^2$ being the quasi-particle-number fluctuations (QNF) at finite temperature and angular momentum
\begin{equation}
\delta{\cal N}_k^2=(\delta{\cal N}_k^+)^2+(\delta{\cal N}_k^-)^2 = n_k^+(1-n_k^+)+n_k^-(1-n_k^-) ~. \label{QNF}
\end{equation}
The coefficients $u_k$ and $v_k$, quasiparticle energies $E_k$, and quasiparticle occupation numbers $n_k^\pm$ are defined as 
\begin{eqnarray}
u_k^2&=&\frac{1}{2}\left(1+\frac{\epsilon_k-Gv_k^2-\lambda}{E_k}\right)~, \hspace{5mm}
v_k^2=\frac{1}{2}\left(1-\frac{\epsilon_k-Gv_k^2-\lambda}{E_k}\right)~, \\
E_k&=&\sqrt{(\epsilon_k-Gv_k^2-\lambda)^2+\Delta_k^2}~, \hspace{5mm}
n_k^{\pm} = \frac{1}{1+e^{\beta(E_k \mp \omega m_k)}}~,\hspace{5mm} \label{uv}
\end{eqnarray}
where $\beta = 1/T$ is the inverse of temperature. It is worth mentioning that the FTBCS1 gap equation (\ref{Gapeq}), which is level dependent, consists of two parts. The first part, $\Delta$, is similar to the conventional FTBCS, and the second part, $\delta\Delta_k$, contains the QNF. By omitting the QNF $\delta{\cal N}_k^2$, one recovers the conventional FTBCS equations from the FTBCS1 ones.

Within the FTBCS (FTBCS1), the total grand-partition function $\Omega$ is given as the sum of the grand-partition functions for protons $\Omega_Z$ and neutrons $\Omega_N$ \cite{Moretto}
\begin{equation}
\Omega = \Omega_N + \Omega_Z = S_N + S_Z + \alpha_N N +\alpha_Z Z + \mu M - \beta{\cal E} ~,
\label{Ome}
\end{equation}
where the total (internal) energy ${\cal E}$ and entropy $S$ are calculated as
\begin{equation}
{\cal E} = \langle H\rangle=\frac{\partial\Omega}{\partial\beta} ~,\label{Etol}
\end{equation}
\begin{equation}
  S = -\sum_k[n_k^+ {\rm ln} n_k^+ + (1-n_k^+){\rm ln}(1-n_k^+)+ n_k^{-} {\rm ln} n_k^{-} + (1-n_k^{-}){\rm ln}(1-n_k^{-})] ~, \label{Stol}
\end{equation}
with $\alpha = \beta\lambda$ and $\mu = \beta\omega$. The level density is calculated based on the inverse Laplace transformation of the grand-partition function (\ref{Ome}). It reads 
\begin{equation}
\rho({\cal E},M) = \frac{e^{(S_N+S_Z)}}{(2\pi)^2\sqrt{D}} ~, \label{Rho1}
\end{equation} 
where the determinant $D$ is given as 
\begin{equation}
D = \left| \begin{array}{cccc}
\frac{\partial^2\Omega}{\partial\alpha_N^2} & \frac{\partial^2\Omega}{\partial\alpha_N\partial\alpha_Z} & \frac{\partial^2\Omega}{\partial\alpha_N\partial\mu} & \frac{\partial^2\Omega}{\partial\alpha_N\partial\beta} \\
\frac{\partial^2\Omega}{\partial\alpha_Z\partial\alpha_N} & \frac{\partial^2\Omega}{\partial\alpha_Z^2} & \frac{\partial^2\Omega}{\partial\alpha_Z\partial\mu} & \frac{\partial^2\Omega}{\partial\alpha_Z\partial\beta} \\
\frac{\partial^2\Omega}{\partial\mu\partial\alpha_N} & \frac{\partial^2\Omega}{\partial\mu\partial\alpha_Z} & \frac{\partial^2\Omega}{\partial\mu^2} & \frac{\partial^2\Omega}{\partial\mu\partial\beta} \\
\frac{\partial^2\Omega}{\partial\beta\partial\alpha_N} & \frac{\partial^2\Omega}{\partial\beta\partial\alpha_Z} & \frac{\partial^2\Omega}{\partial\beta\partial\mu} & \frac{\partial^2\Omega}{\partial\beta^2}
\end{array} \right | ~. \label{Det}
\end{equation}
The total NLD $\rho({\cal E})$ is calculated based on the sum of all $J$-dependent NLD $\rho({\cal E}) = \sum_{J}(2J+1)\rho({\cal E},J)$ \cite{Gilbert}, where $\rho({\cal E},J)$ is obtained by differentiating $\rho({\cal E},M)$, namely $\rho({\cal E},J)=\rho({\cal E},M=J) - \rho({\cal E},M=J+1)$ \cite{Ericson}.
\section{Results}
\begin{figure}[h]
\begin{center}
\includegraphics[width=9 cm]{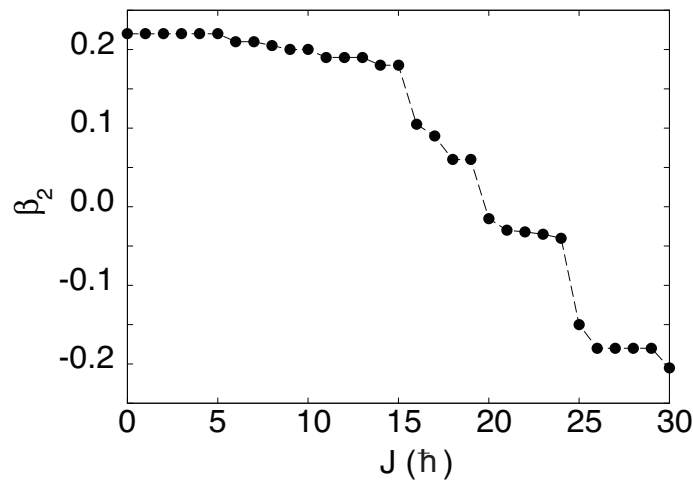}
\caption{\label{Fig1} Quadrupole deformation parameter $\beta_2$ of the Woods-Saxon potential as functions of the total angular momentum $J$ obtained within the FTBCS (FTBCS1).}
\end{center}
\end{figure}

The numerical calculations are carried out for $^{104}$Pd nucleus, whose single-particle spectra are taken from the axially deformed Woods-Saxon potential including the spin-orbit and Coulomb interactions \cite{Cwiok}. The pairing interaction parameters $G_{N,Z}$ are adjusted so that the neutron and proton gaps obtained within the FTBCS (FTBCS1) at $T $ = 0 match those given by the empirical odd-even mass differences \cite{Ring}. The quadrupole deformation parameters $\beta_2$ of the Woods-Saxon potential are adjusted so that the NLD obtained at different values of $J$ fit best the empirical ones, which are used in the CASCADE code to fit the experimental proton spectrum~\cite{Mitra}, especially in the region where the enhancement of NLD is observed. The variation of $\beta_2$ with $J$ is plotted in Fig. \ref{Fig1}. This figure clearly shows that $^{104}$Pd nucleus undergoes a shape transition from the prolate shape ($\beta_2 > 0$) to the oblate one ($\beta_2 < 0$) at around $J$ = 20 $\hbar$. This transition seems to be reasonable in this mass region because of an alignment of protons in $g_{9/2}$ and neutrons in $h_{11/2}$ orbits\cite{ShapeTrans}.

\begin{figure}[h]
\begin{center}
\includegraphics[width=11 cm]{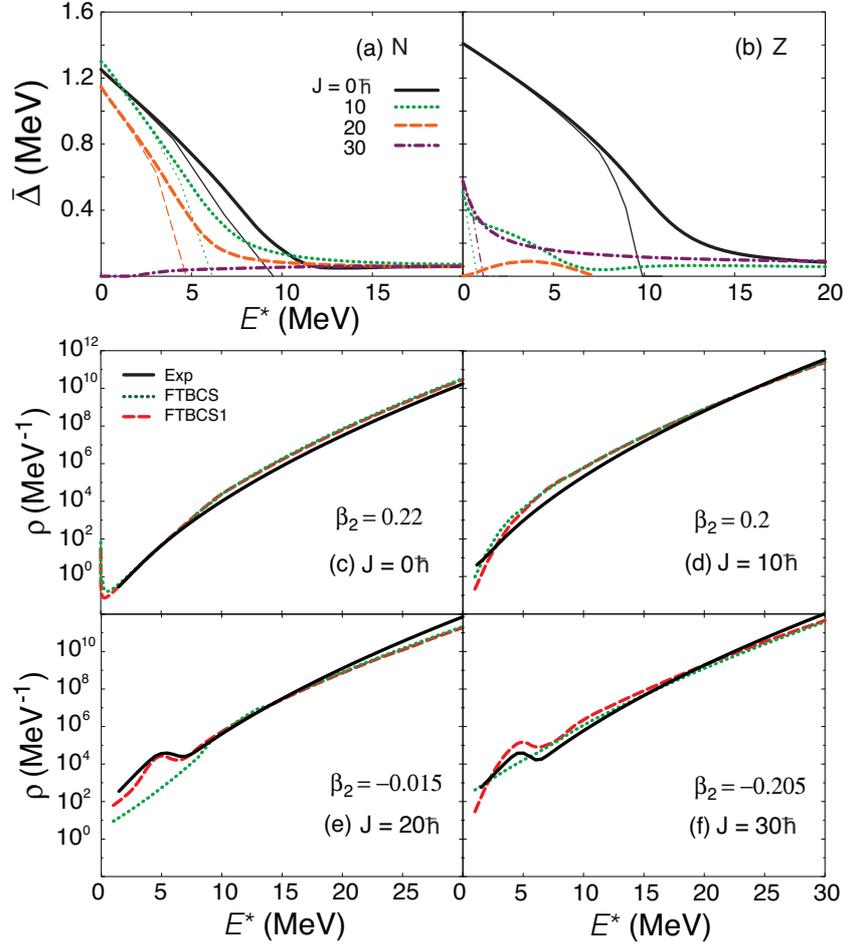}
\caption{\label{Fig2} Level-weighted pairing gaps $\bar\Delta\equiv\sum_k\Delta_k/N_L$ with $N_L$being the number of single-particle levels $k$ for neutron ($N$) (a) and protons ($Z$) (b) and total NLD (c) - (f) as function of excitation energy $E^*$ obtained within the FTBCS and FTBCS1 at different values of $J$ and $\beta_2 = 0.276$. The thin and thick lines in (a) and (b) denote the FTBCS and FTBCS1 results, respectively, whereas the dotted and dashed lines in (c) - (f) respectively stand for the FTBCS and FTBCS1 total level densities. The solid lines in (c) - (f) are the empirical NLD employed in the CASCADE code to fit the experimentally extracted proton spectra in the reaction $^{12}$C + $^{93}$Nb $\rightarrow$ $^{105}$Ag$^*$ $\rightarrow$ $^{104}$Pd$^* + p$ in Ref. \cite{Mitra}.}
\end{center}
\end{figure}

Shown in Figs. \ref{Fig2} (a)-(f) are the pairing gaps and NLD as functions of excitation energy. The latter is defined as $E^* = {\cal E}(T, M) - {\cal E}(0, M)$, which are obtained within the FTBCS and FTBCS1 at different values of $J$ and $\beta_2$. It is clear from Figs. \ref{Fig2} (a) and (b) that the neutron and proton gaps, which are obtained within the FTBCS (thin lines), decrease with increasing $E^*$ at all $J$, and collapse at some critical values $E_c^*$. At the same time, because of the QNF, the pairing gaps, predicted by the FTBCS1, do not collapse but monotonically decrease with increasing $E^{*}$, and remain finite even at $E^*=$ 20 MeV, except the proton gap at $J$ = 20$\hbar$ [dashed lines in Fig. \ref{Fig2} (b)]. The latter is zero at $E^*$ = 0, increases with increasing $E^*$ to reach a maximum at $E^{*}\approx$ 3 MeV, and then decreases to vanish at $E^* \approx$ 7 MeV. This feature is caused by the change of shell structure, which takes place in the shape transition from prolate to oblate at this $J$ value. The pairing reentrance is therefore present in the FTBCS1 gaps at $J$ = 20$\hbar$ for protons and $J$ = 30$\hbar$ for neutrons [dash dotted lines in Fig. \ref{Fig2} (a)], whereas no signature of this effect is seen in the pairing gaps obtained within the FTBCS. As the result, there is no enhancement of the NLD obtained within the FTBCS [dotted lines in Fig. \ref{Fig2} (c) - (f)], whereas two local enhancements are seen in the FTBCS1 NLD in good agreement with the empirical ones at exactly two values of $J =$ 20 and 30$\hbar$, where the pairing reentrance takes place, 

\section{Conclusions}

This work studies the pairing reentrance phenomenon in a warm rotating $^{104}$Pd nucleus by the analyzing the pairing gaps and NLD obtained within the FTBCS and FTBCS1 theories including finite angular momentum. The results obtained show that the pairing reentrance takes place only in the pairing gaps obtained within the FTBCS1 (e.g., for protons at $J = 20\hbar$ and neutrons at $J = 30\hbar$), whereas this effect does not appear in the FTBCS gaps. This leads to the local enhancements of the NLD obtained within the FTBCS1 at low excitation energy and high angular momentum in agreement with the empirical NLD. This agreement indicates that the observed enhancement of the NLD might be the first experimental detection of the pairing reentrance in a finite nucleus.

The numerical calculations were carried out using the Integrated Cluster of Clusters (RICC) system at RIKEN. N.Q.H. thanks for the support of the National Foundation for Science and Technology Development (NAFOSTED) of Vietnam through Grant No.103.04-2013.08.

\end{document}